\documentclass[fleqn,usenatbib]{mnras}

\usepackage{newtxtext,newtxmath}

\usepackage[T1]{fontenc}
\usepackage{ae,aecompl}

\usepackage{graphicx}	
\usepackage{amsmath}	
\usepackage{amssymb}	
\usepackage{adjustbox}
\usepackage{subcaption} 
\usepackage{threeparttable}

\title[First detection of HOCH$_2$CN in the ISM]{First detection of the pre-biotic molecule glycolonitrile (HOCH$_2$CN) in the interstellar medium}

\author[S.Zeng et al.]{
S. Zeng$^{1}$,\thanks{E-mail: s.zeng@qmul.ac.uk}
D. Qu\'enard$^{1}$,
I. Jim\'enez-Serra$^{2}$,
J. Mart\'in-Pintado$^{2}$,
V. M. Rivilla$^{3}$,
\newauthor{
L. Testi$^{3,4,5}$, and R. Mart\'in-Dom\'enech$^{6}$}
\\
$^{1}$School of Physics and Astronomy, Queen Mary University of London, Mile End Road, E1 4NS London, United Kingdom\\
$^{2}$Centro de Astrobiolog\'ia (CSIC, INTA), Ctra. de Ajalvir, km. 4, Torrej\'on de Ardoz, 28850 Madrid, Spain \\
$^{3}$INAF-Osservatorio Astrofisico di Arcetri, Largo Enrico Fermi 5, I-50125, Florence, Italy\\
$^{4}$Excellence Cluster "Universe", Boltzmann str. 2, D-85748 Garching bei Muenchen, Germany \\
$^{5}$ESO/European Southern Observatory, Karl Schwarzschild str. 2, D-85748, Garching, Germany \\
$^{6}$Harvard-Smithsonian Center for Astrophysics, 60 Garden Street, Cambridge, MA 02138, USA
\\
}

\date{Accepted XXX. Received YYY; in original form ZZZ}

\pubyear{2018}

\begin{document}
\label{firstpage}
\pagerange{\pageref{firstpage}--\pageref{lastpage}}
\maketitle

\begin{abstract} 
Theories of a pre-RNA world suggest that glycolonitrile (HOCH$_2$CN) is a key species in the process of ribonucleotide assembly, which is considered as a molecular precursor of nucleic acids. In this Letter, we report the first detection of this pre-biotic molecule in the interstellar medium (ISM) by using ALMA data obtained at frequencies between 86.5$\,$GHz and 266.5$\,$GHz toward the Solar-type protostar IRAS16293-2422 B. A total of 15 unblended transitions of HOCH$_2$CN were identified. Our analysis indicates the presence of a cold (T$\rm _{ex}$=24$\pm$8$\,$K) and a warm (T$\rm _{ex}$=158$\pm$38$\,$K) component meaning that this molecule is present in both the inner hot corino and the outer cold envelope of IRAS16293 B. The relative abundance with respect to H$_2$ is (6.5$\pm$0.6)$\times$10$^{-11}$ and $\geq$(6$\pm$2)$\times$10$^{-10}$ for the warm and cold components respectively. Our chemical modelling seems to underproduce the observed abundance for both the warm and cold component under various values of the cosmic-ray ionisation rate ($\zeta$). Key gas phase routes for the formation of this molecule might be missing in our chemical network.  
\end{abstract}




\section{Introduction}
Nucleobases are nitrogen heterocycles that are key components in biological nucleic acids. Theories of a primordial RNA world suggest that these nitrogen heterocycles could have been synthesized easily from hydrogen cyanide \citep[or HCN; see][]{oro1961}. In this chemical scheme, one of the key precursors towards adenine formation (one of the two two-ring N-heterocycle nucleobases known) is glycolonitrile \citep[HOCH$_2$CN;][]{menor-salvan2012}\footnote{Glycolonitrile is also known as hydroxyacetonitrile.}. Indeed, it has been shown that the freezing of dilute solutions of HOCH$_2$CN not only produces adenine \citep{schwartz1982} but it also strongly accelerates HCN oligomerization \citep{schwartz1982b}. HOCH$_2$CN could even be a precursor of glycine if reacting with ammonia in eutectic (ice-water) solutions (Menor-Salv\'an, priv. comm.).

In the ISM, HOCH$_2$CN could form on icy grain surfaces via the reaction between formaldehyde (H$_2$CO) and hydrogen cyanide \citep[HCN;][]{danger_hydroxyacetonitrile_2012}. \citet[][]{danger_hydroxyacetonitrile_2013} later on showed that its photo-destruction is expected to yield species such as cyanogen (NCCN), formylcyanide (CHOCN), and ketenimine (CH$_2$CNH). The latter two have been detected toward the hot core Sgr B2(N-LMH) \citep{lovas_detection_2006,remijan_detection_2008}.
As a structural isomer of HOCH$_2$CN, significant amounts of methyl isocyanate (CH$_3$NCO) have been readily measured in astrophysical environments such as the high-mass hot cores Sgr B2(N) and Orion KL \citep{halfen_interstellar_2015,belloche_rotational_2017,cernicharo_rigorous_2016}, and the Solar-type protostar IRAS16293-2422 \citep[I16293 hereafter;][]{ligterink_alma-pils_2017,martin-domenech_detection_2017}. However detection of HOCH$_2$CN remains to be reported.

I16293 is a well-studied Class 0 protostar located in the $\rho$ Ophiuchi star-forming region, at a distance of 141$^{+31}_{-21}$ pc \citep{Dzib2018}. It has a cold outer envelope (with spatial scales of up to $\sim$6000 au) \citep{Jaber2017} and a hot corino at scales of $\sim$100 au \citep{jorgensen_alma_2016}. Due to its hot-core-like properties, a wealth of complex organic molecules (COMs) have been reported toward its two binary components: I16293A and I16293B, separated by 5$^{\prime\prime}$ in the plane of sky \citep{wootten1989,looney2000}. Here, we report the first detection of HOCH$_2$CN towards I16293B at frequencies $\leq$270$\,$GHz using the Atacama Large Millimeter Array (ALMA). 

\section{Observations}
We searched for HOCH$_2$CN using several publicly available datasets in Bands 3, 4, and 6 in the ALMA archive (projects ID: $\#$2011.0.00007.SV, $\#$2012.1.00712.S, $\#$2013.1.00061.S, $\#$2013.1.00352.S, and $\#$2015.1.01193.S \citep[e.g.][]{jorgensen_alma_2016,ligterink_alma-pils_2017,martin-domenech_detection_2017}). By excluding datasets in Bands 7 and 8, we limit the spectral confusion seen at high frequencies, which facilitates the continuum subtraction performed in the uv-plane before imaging using line-free channels from the observed spectra. In addition, the line optical depth effect by the dust can also be avoided since the continuum emission toward the innermost parts of I16293 B is likely optically thick at frequencies >300 GHz \citep{zapata2013,jorgensen_alma_2016}. As in \citet{martin-domenech_detection_2017}, the standard ALMA calibration scripts and the Common Astronomy Software Applications package were used for data calibration and imaging. 

Our dataset covers a total bandwidth of $\sim$7$\,$GHz spread in multiple spectral ranges between 86.5$\,$GHz and 266.5$\,$GHz. The beam sizes range between 1.42$^{\prime\prime}$ and 1.85$^{\prime\prime}$ with spectral resolutions of 61-282$\,$kHz that correspond to velocity resolutions of 0.1-0.9$\,$km s$^{-1}$. To facilitate the analysis, a circle of 1.6$^{\prime\prime}$ diameter was used to extract the spectra from each dataset at the position of I16293 B: $\alpha$(J2000.0)=16$^h$32$^m$22.61$^s$, $\delta$(J2000.0)=-24$^{\circ}$28$^{\prime}$32.44$^{\prime\prime}$. Note that the beam sizes are much larger than the source size (0.5$^{\prime\prime}$) \citep{jorgensen_alma_2016,lykke_alma-pils_2017,martin-domenech_detection_2017} and hence the flux measured within our circular support of 1.6$^{\prime\prime}$ diameter contains all the flux independently of the beam. Angular resolutions $\leq$1.9$^{\prime\prime}$ are sufficient to resolve source B from source A in the I16293 binary, and thus the emission from I16293 B exhibits line profiles $\leq$2 kms$^{-1}$.

\section{Results}
\begin{figure*}
\includegraphics[width = 0.8\textwidth]{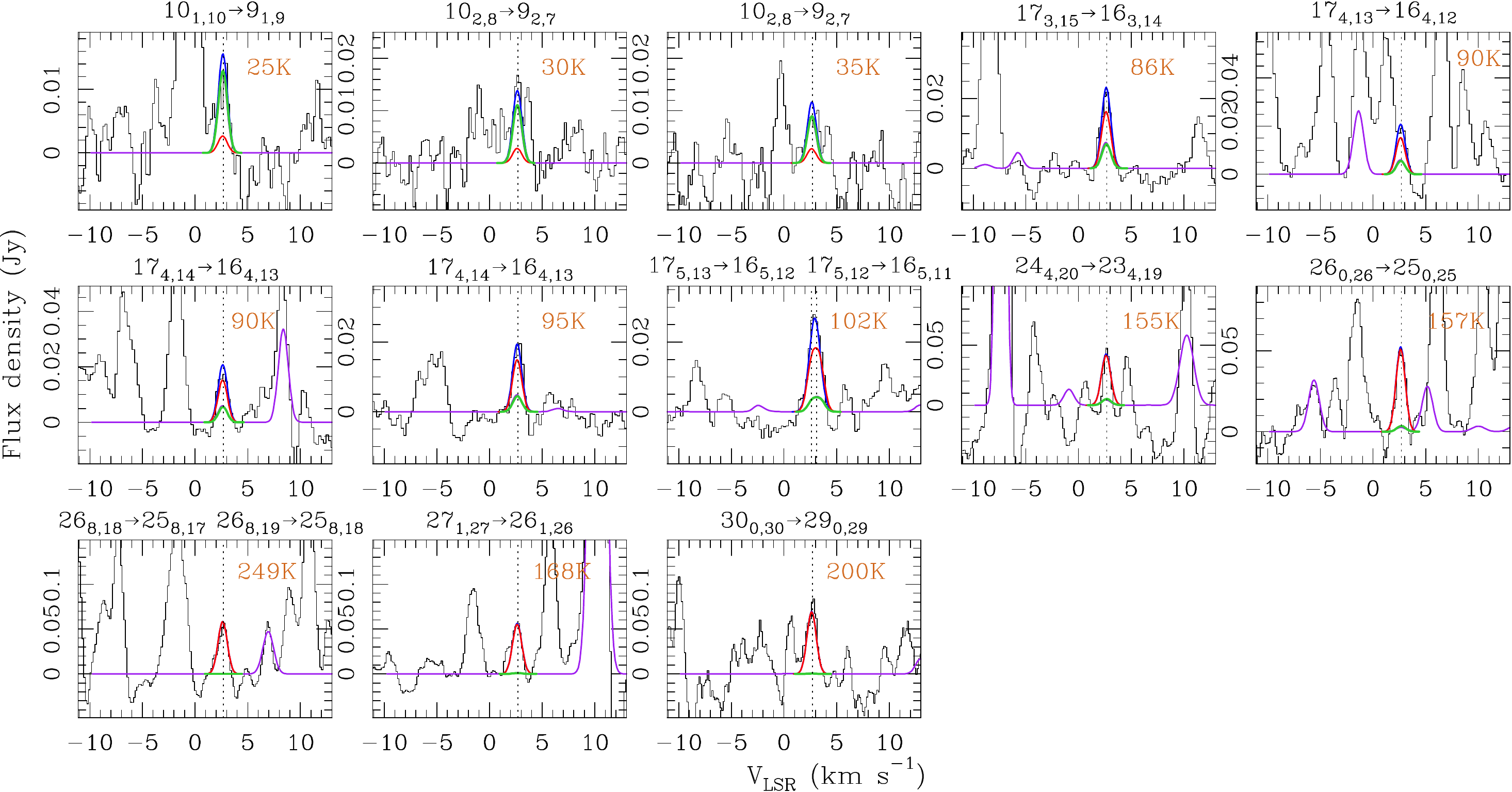}
\vspace{-2.5mm}
 \caption{Spectra of all unblended transitions of HOCH$_2$CN detected toward I16293B. Black line indicates the observed spectra. Dashed line indicates each transition centred at $V_{\rm LSR}$=2.6$\,$km s$^{-1}$. Purple lines indicates emission arise from other species. Blue line indicates the best fit considering both temperature components. Red and green line indicates the best fit for $T_{\rm ex}$=158$\pm$38$\,$K and $T_{\rm ex}$=24$\pm$8$\,$K respectively. The $E_{\rm up}$ value for each line is shown in brown in the right upper part of each panel whilst the quantum number J, K$_a$ and K$_c$ for each transition is shown above each panel.}
    \label{fig:line-profiles}
\end{figure*}

\begin{figure}
\centering
\includegraphics[width =0.45\textwidth]{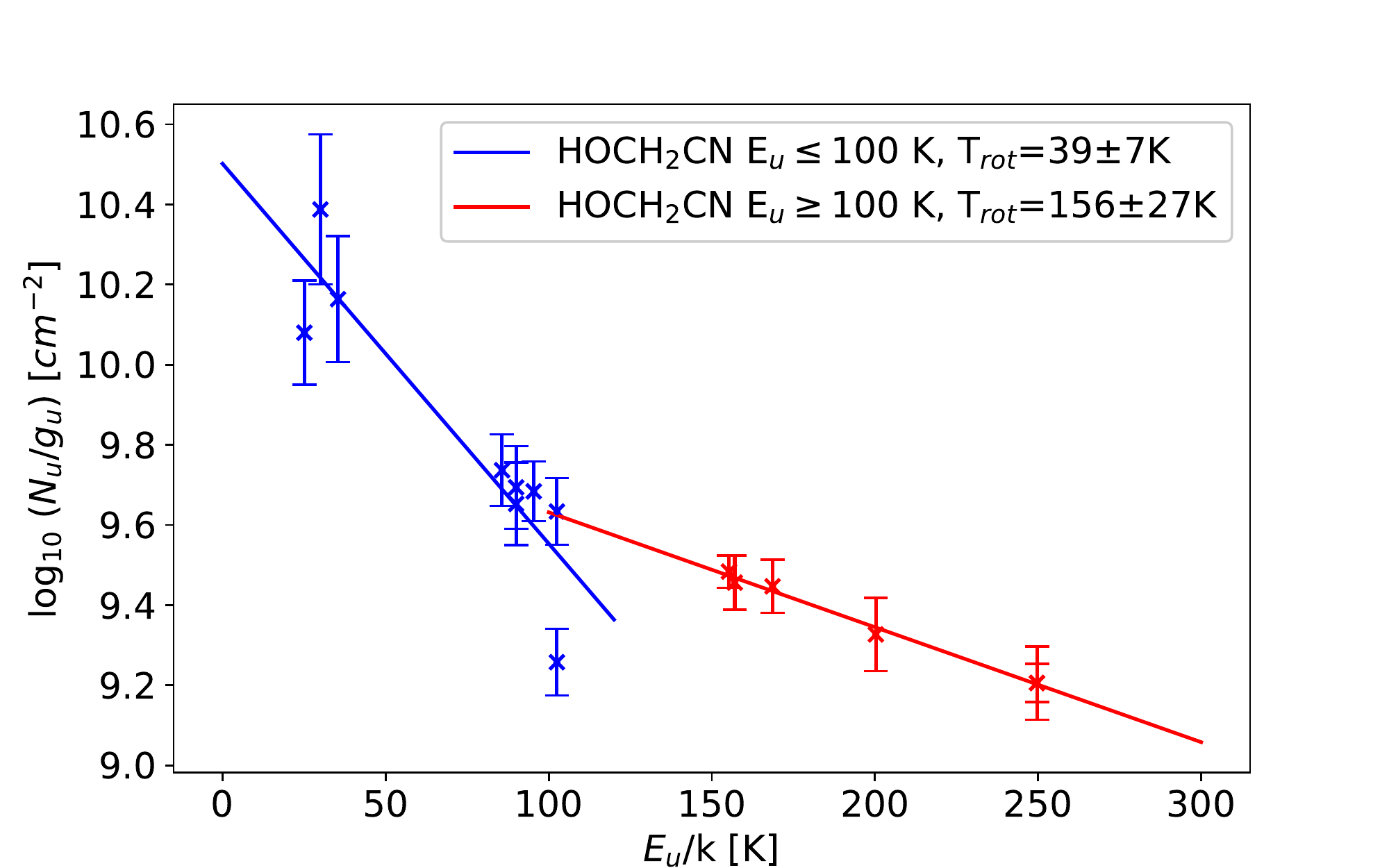}
\vspace{-2.5mm}
 \caption{Rotational diagram of HOCH$_2$CN obtained assuming optically thin emission. Two temperature components are needed to fit the data. Blue x indicate transitions with $E_{\rm up}\leq$100$\,$K (probing the cold component), while orange x indicate transitions with $E_{\rm up}$$\geq$100$\,$K (from the warm component). Note that in this rotational diagram we are assuming an extended source size for all transitions. The final value of $N_{\rm tot}$ for the warm component has to be corrected by the source size of the hot corino of 0.5$"$ (i.e. by applying a correction factor of $\frac{(0.5^{\prime\prime})^2+(1.6^{\prime\prime})^2}{(0.5^{\prime\prime})^2}$=11.2).}
    \label{fig:rotational-diagram}
\end{figure}

The line identification and analysis were performed with \textsc{madcuba} package\footnote{Madrid Data Cube Analysis on ImageJ is a software developed at the Astrobiology Center in Madrid \citep[see][]{rivilla2016}.}. For HOCH$_2$CN, we have used the spectroscopic parameters from the Cologne Database for Molecular Spectroscopy (CDMS)\footnote{http://www.astro.uni-koeln.de/cdms} \citep{muller_cologne_2001,muller_cologne_2005,endres_cologne_2016} based on the laboratory work of \citet{margules_submillimeter_2017}. Within the frequency range covered, we have identified 35 transitions with peak fluxes $\geq$3$\sigma$, where $\sigma$ is the rms noise level measured in the spectra. 15 out of 35 transitions are unblended from other molecular species whilst the rest are blended but show consistency with the observed spectra. Spectroscopic information of the unblended transitions are summarized in Table \ref{tab:detected-transitions}. The \textsc{madcuba-slim} package was used to produce synthetic spectra for each line profile by considering Local Thermodynamical Equilibrium (LTE) and line opacity effects. For the typical densities and linear scales imaged with ALMA toward I16293, LTE is a good approximation \citep[see][]{jorgensen_alma_2016}. In addition, no collisional coefficients are available for HOCH$_2$CN. The parameters column density ($N_{\rm tot}$), excitation temperature ($T_{\rm ex}$), velocity ($V_{\rm LSR}$), and linewidth ($\Delta \nu$) were adjusted manually to match the synthetic spectra to the line profiles. Then the \textsc{madcuba-autofit} tool was employed to provide the best non-linear least-squared fit using the Levenberg-Marquardt algorithm. Since a single temperature was unable to reproduce all the observed spectra, two temperature components were invoked, which is also agreed with our rotational diagram analysis (see Figure\,\ref{fig:line-profiles} and \ref{fig:rotational-diagram}). However we cannot rule out the possibility that LTE might not apply which may explain why a single temperature cannot reproduce the observations. While the low-E$_{\rm up}$ lines require a prominent contribution from a cold component (with $T_{\rm ex}$=24$\pm$8$\,$K; see green lines), the high-E$_{\rm up}$ lines are dominated by a warm component (with $T_{\rm ex}$=158$\pm$38$\,$K; red lines). The warm component is best fitted with $V_{\rm LSR}$=2.62$\pm$0.03$\,$km s$^{-1}$ and $\Delta \nu$=1.03$\pm$0.03$\,$km s$^{-1}$, whilst the cold component is best fitted with $V_{\rm LSR}$=2.62$\pm$0.15$\,$km s$^{-1}$ and with $\Delta \nu$ fixed to 1$\,$km s$^{-1}$. The line parameters derived for warm component are consistent with those measured for other molecules in I16293B \citep{jorgensen_alma_2016, martin-domenech_detection_2017, ligterink_alma-pils_2017,rivilla_first_2018}. As in the analysis of CH$_3$NCO by \citet{martin-domenech_detection_2017} toward I16293B, the source size was fixed to 0.5$^{\prime\prime}$ for the warm component but an extended source size (>10000 au or $>$20$^{\prime\prime}$) was considered for the cold component. Our analysis is not affected by line optical depth effects since all HOCH$_2$CN transitions are optically thin ($\tau \leq$0.02). 

\begin{table}
 \centering
 \Huge
 \caption{Unblended transitions of HOCH$_2$CN identified toward I16293B with ALMA.}
\begin{adjustbox}{width=0.5\textwidth}
\begin{tabular}{cccccccc}
\hline
\hline
Frequency & Transition & log$A_{\rm ul}$ & $E_{\rm up}$ & Area & $\Delta \nu$ & rms\\
(GHz) & (J, $K_{\rm a}$, $K_{\rm c}$) &  & (K) & (mJy km s$^{-1}$) & (km s$^{-1}$) & (mJy km s$^{-1}$) \\
\hline
89.650064 & (10,1,10)-(9,1,9),v=0-0 & -4.67 & 25 & 17$\pm$4 & 1.1$\pm$0.4 & 2\\ 
92.831280 & (10,2,8)-(9,2,7),v=1-1 & -4.63 & 35 & 13$\pm$3 & 1.5$\pm$0.3 & 2\\
92.898070 & (10,2,8)-(9,2,7),v=0-0 & -4.63 & 30 & 15$\pm$4 & 1.0$\pm$0.3 & 2\\
156.881298 & (17,4,14)-(16,4,13),v=1-1 & -3.95 & 95 & 21$\pm$5 & 1.0$\pm$0.1 & 2\\
156.907235 & (17,5,13)-(16,5,12),v=0-0 & -3.97 & 102 & 19$\pm$5 & 0.9$\pm$0.2 & 2\\
156.907641 & (17,5,12)-(16,5,11),v=0-0 & -3.97 & 102 & 19$\pm$5 & 0.9$\pm$0.1 & 2\\
156.917954 & (17,3,15)-(16,3,14),v=1-1 & -3.93 & 86 & 25$\pm$7 & 1.1$\pm$0.1 & 2\\
156.995601 & (17,4,14)-(16,4,13),v=0-0 & -3.96 & 90 & 23$\pm$6 & 1.1$\pm$0.1 & 2\\
157.015484 & (17,4,13)-(16,4,12),v=0-0 & -3.79 & 90 & 23$\pm$6 & 0.9$\pm$0.4 & 2\\
222.111772 & (24,4,20)-(23,4,19),v=0-0 & -3.48 & 155 & 47$\pm$8 & 1.0$\pm$0.2 & 3\\
231.394834 & (26,0,26)-(25,0,25),v=1-1 & -3.42 & 157 & 58$\pm$9 & 0.9$\pm$0.1 & 4\\
239.783676 & (26,8,19)-(25,8,18),v=1-1 & -3.41 & 249 & 32$\pm$3 & 0.9$\pm$0.1 & 3\\
239.783676 & (26,8,18)-(25,8,17),v=1-1 & -3.41 & 249 & 32$\pm$3 & 0.9$\pm$0.1 & 3\\
239.752553 & (27,1,27)-(26,1,26),v=1-1 & -3.31 & 168 & 62$\pm$9 & 0.9$\pm$0.1 & 3\\
266.232097 & (30,0,30)-(29,0,29),v=0-0 & -3.21 & 200 & 77$\pm$9 & 1.0$\pm$0.1 & 3\\
\hline
\hline
\end{tabular}
\end{adjustbox}
\label{tab:detected-transitions}
\end{table}



To strengthen the need for two temperatures, we built the rotational diagram using all unblended HOCH$_2$CN lines and assuming optically thin emissions (Figure \ref{fig:rotational-diagram}). Two linear least square fits are required to fit simultaneously the transitions with E$\rm_{up}$$\leq$100$\,$K and $E_{\rm up}$$\geq$100$\,$K. The two rotational temperature $T_{\rm rot}$ inferred from the rotational diagram are 39$\pm$7$\,$K and 156$\pm$27$\,$K. They are consistent (within the errors) with the $T_{\rm ex}$ computed by \textsc{madcuba} of 24$\pm$8$\,$K and 158$\pm$38$\,$K for the cold and warm components respectively. The inferred column density is $N_{\rm tot}$=(8.5$\pm$2.6)$\times$10$^{13}$$\,$cm$^{-2}$ and $N_{\rm tot}$=(1.8$\pm$0.1)$\times$10$^{15}$$\,$cm$^{-2}$ for the cold and warm components respectively. An H$_2$ column density of N(H$_2$)=2.8$\times$10$^{25}$ cm$^{-2}$ was adopted from \citep{martin-domenech_detection_2017}, yielding an HOCH$_2$CN abundance of (6.5$\pm$0.6)$\times$10$^{-11}$ for the warm component. Given that the cold component of HOCH$_2$CN is likely to trace the outer envelope, an H$_2$ column density of N(H$_2$)=1.5$\times$10$^{23}$ cm$^{-2}$ was adopted using the envelope profile constrained by \citet{crimier2010}. The resulting HOCH$_2$CN abundance for the cold component is therefore (6$\pm$2)$\times$10$^{-10}$. Since this N(H$_2$) refers to the whole envelope, our abundance must be considered as a lower limit. 

We note that \citet{Pineda2012} also needed to invoke the presence of two-temperature components to reproduce the moderately optically-thick lines of CH$_3$OCHO and H$_2$CCO toward I16293B. However, their inferred temperatures are T=3$\,$K and T=45-60$\,$K for the cold and warm components respectively, i.e. very different from the ones derived in this work. Given the complex structure of the source with known temperature/density and kinematic gradients, different COMs likely probe different temperature/density regions within the envelope. This may explain why the HOCH$_2$CN lines do not show any shift in V$_{\rm lsr}$ between the cold and warm components, or any broadening for the high-$E_{\rm up}$ lines, as expected from the infall motions reported in \citet{Pineda2012}. In addition, the HOCH$_2$CN lines are optically thin ($\tau$$<$0.02) with higher excitation temperatures, making very unlikely the presence of inverse P-Cygni profiles.


As studied by \citet{danger_hydroxyacetonitrile_2012}, HOCH$_2$CN can form in concurrence with aminomethanol (NH$_2$CH$_2$OH), a precursor of aminoacetonitrile (NH$_2$CH$_2$CN) and, possibly, of glycine (NH$_2$CH$_2$COOH) \citep{Belloche2008}. The yielding ratio between NH$_2$CH$_2$OH and HOCH$_2$CN is sensitive to the initial ratio between NH$_3$ and ammonium salt ([NH$_4$$^{+-}$CN]) in the ice (see Figure 4 in their paper). In I16293B, we are unable to search for NH$_2$CH$_2$OH due to the lack of laboratory or theoretical line frequencies and strengths, but its daughter molecule aminoacetonitrile (NH$_2$CH$_2$CN) is undetected in the warm component with a 3$\sigma$ upper limit $N_{\rm tot}$ $\leq$1.7$\times$10$^{14}$$\,$cm$^{-2}$ (or an H$_2$ relative abundance of $\leq$6.2$\times$10$^{-12}$). This could suggest that in I16293B the formation of HOCH$_2$CN is favoured over NH$_2$CH$_2$OH. The photo-products of HOCH$_2$CN, CHOCN and CH$_2$CNH for which spectroscopic data exist \citep{danger_hydroxyacetonitrile_2012}, are also undetected in the warm component with 3$\sigma$ abundance upper limits $\leq$1.3$\times$10$^{-11}$ and $\leq$7.5$\times$10$^{-12}$ respectively (i.e. $\sim$6 to 11 times less abundant than HOCH$_2$CN).

\begin{table}
\centering
\Large
\caption{Gas and grain surface chemical network of HOCH$_2$CN.}
\begin{adjustbox}{width=0.5\textwidth}
\begin{tabular}{lcccc}
\hline
\hline
& Chemical reaction											&	$\alpha$		&	$\beta$	&	$\gamma$\\
\hline
1 & $\rm \#H_2CO + \#NH_3 \rightarrow \#HOCH_2NH_2$				&	$0.50(-02)$	&	$0$		&	$529.2$\\
2& $\rm \#H_2CO + \#CN \rightarrow \#HOCH_2CN$					&	$2.80(-01)$	&	$0$		&	$457.0$\\
3& $\rm \#HOCH_2CN + PHOTON \rightarrow \#CH_2CNH + \#OH$		&	$2.83(-17)$	&	$0$		&	$0$\\
4& $\rm \#HOCH_2CN + PHOTON \rightarrow \#H_2CO + \#HCN$		&	$2.12(-16)$	&	$0$		&	$0$\\
5& $\rm \#HOCH_2CN + PHOTON \rightarrow \#CHOCN + \#H + \#H$		&	$4.24(-17)$	&	$0$		&	$0$\\
6& $\rm \#HOCH_2NH_2 + PHOTON \rightarrow \#NH_2CHO + \#H_2$	&	$2.19(-15)$	&	$0$		&	$0$\\
7& $\rm \#H_2CO + PHOTON \rightarrow \#CO + \#H_2$				&	$4.88(-16)$	&	$0$		&	$0$\\
8& $\rm \#NH_3 + PHOTON \rightarrow \#NH_2 + \#H$					&	$1.56(-16)$	&	$0$		&	$0$\\
\hline
9& $\rm H_3^+ + HOCH_2CN \rightarrow CH_3OH + CN + H_2$			&	$2.50(-10)$	&	$-0.5$	&	$0$\\
10& $\rm H_3^+ + HOCH_2CN \rightarrow CH_3CN + OH + H_2$			&	$2.50(-10)$	&	$-0.5$	&	$0$\\
11& $\rm H_3^+ + HOCH_2CN \rightarrow CH_2OH + HCN + H_2$			&	$2.50(-10)$	&	$-0.5$	&	$0$\\
12& $\rm H_3^+ + HOCH_2CN \rightarrow CH_2CN + H_2O + H_2$		&	$2.50(-10)$	&	$-0.5$	&	$0$\\
13& $\rm HCO^+ + HOCH_2CN \rightarrow CH_3OH + CN + CO$			&	$2.73(-10)$	&	$-0.5$	&	$0$\\
14& $\rm HCO^+ + HOCH_2CN \rightarrow CH_3CN + OH + CO$			&	$2.73(-10)$	&	$-0.5$	&	$0$\\
15& $\rm HCO^+ + HOCH_2CN \rightarrow CH_2OH + HCN + CO$		&	$2.73(-10)$	&	$-0.5$	&	$0$\\
16& $\rm HCO^+ + HOCH_2CN \rightarrow CH_2CN + H_2O + CO$		&	$2.73(-10)$	&	$-0.5$	&	$0$\\
17& $\rm H_3O^+ + HOCH_2CN \rightarrow CH_3OH + CN + H_2O$		&	$2.73(-10)$	&	$-0.5$	&	$0$\\
18& $\rm H_3O^+ + HOCH_2CN \rightarrow CH_3CN + OH + H_2O$		&	$2.73(-10)$	&	$-0.5$	&	$0$\\
19& $\rm H_3O^+ + HOCH_2CN \rightarrow CH_2OH + HCN + H_2O$		&	$2.73(-10)$	&	$-0.5$	&	$0$\\
20& $\rm H_3O^+ + HOCH_2CN \rightarrow CH_2CN + H_2O + H_2O$		&	$2.73(-10)$	&	$-0.5$	&	$0$\\
21& $\rm HOCH_2CN + CRPHOT \rightarrow CH_2OH + CN$				&	$6.50(-18)$	&	$0$		&	$2000$\\
22& $\rm HOCH_2CN + CRPHOT \rightarrow CH_2CN + OH$				&	$6.50(-18)$	&	$0$		&	$2000$\\
23& $\rm HOCH_2CN + PHOTON \rightarrow CH_2OH + CN$				&	$2.50(-10)$	&	$0$		&	$1.7$\\
24& $\rm HOCH_2CN + PHOTON \rightarrow CH_2CN + OH$				&	$2.50(-10)$	&	$0$		&	$1.7$\\
\hline
\end{tabular}
\end{adjustbox}
\label{table:network}
\begin{tablenotes}
\vspace{-1mm}
\footnotesize
\item \textbf{Notes.} "$\#$" means that the species is on dust grains. $a(b)$ means $a\times10^b$. Bimolecular rate coefficients are tabulated as $k(T) = \alpha\left(\frac{T}{300}\right)^\beta\exp\left({-\frac{\gamma}{T}}\right)$ in units of cm$^3$\,s$^{-1}$. Rates for reactions 1-2 are from \citet{danger_hydroxyacetonitrile_2012}, while those for reactions 3-6 are from \citet{danger_hydroxyacetonitrile_2013}. Rate coefficients for photo-dissociation (PHOTON) are tabulated as $k = \alpha\exp\left(-\gamma{\rm A_V}\right)\left(\frac{G_0}{\rm 1\,Habing}\right)$ in units of s$^{-1}$. Rate coefficients for cosmic-ray-induced secondary photo-dissociation (CRPHOT) are tabulated as $k = 2\,\alpha\,\gamma\,\zeta\left(\frac{T}{300}\right)^\beta$ in units of s$^{-1}$, with $\zeta$ the cosmic-ray ionisation rate. See \citet{mcelroy2013} and \citet{holdship2017} for more details.
\end{tablenotes}
\end{table}

\section{Chemical modelling}

To reproduce the observed abundance of HOCH$_2$CN in the warm and cold components in I16293B, we have used the gas-grain chemical code \textsc{uclchem}\footnote{http://uclchem.github.io} \citep{viti2004, holdship2017, quenard2018a}. The gas-phase reactions are taken from the UMIST database \citep{mcelroy2013} with additional reactions from \citet{quenard2018a} and \citet{majumdar2017}. The code considers thermal desorption of the ices, as well as grain surface non-thermal processes such as diffusion via thermal hopping, quantum tunnelling \citep{hasegawa1992}, cosmic-rays desorption, direct-UV and secondary-UV desorption \citep{mcelroy2013, holdship2017}, and chemical reactive desorption \citep{minissale2016}.

For HOCH$_2$CN, we have updated the chemical network of \citet{quenard2018a} with the laboratory experiments on grain surfaces of \citet{danger_hydroxyacetonitrile_2012,danger_hydroxyacetonitrile_2013}. We have also added several ion-neutral destruction reactions involving HOCH$_2$CN and three major ions: H$_3^+$, HCO$^+$ and H$_3$O$^+$ to the gas phase network. To our knowledge, their reaction rates are unknown but, at first approximation, we assume that they are similar to those involving CH$_3$NCO and keep the different branching ratios \citep[see][]{majumdar2017}. No gas phase formation reactions of HOCH$_2$CN have been added to the network due to the lack of either laboratory experiments or theoretical calculations. The list of reactions added to the gas and grain surface chemical network is shown in Table \ref{table:network}. The network contains 372 species (247 in the gas phase and 125 on the grain surface) and 3525 reactions. The starting elemental abundances are the same as in \citet{quenard2018a}. Binding energies of molecules are taken from \citet{wakelam2017} and the binding energy for HOCH$_2$CN (of 6980\,K) is from \citet[][]{danger_hydroxyacetonitrile_2012}. 

The chemical modelling is performed in three steps. The first step corresponds to the \textit{diffuse cloud phase}, we followed the chemistry of a low-density cloud (n$_{\rm H}=10^2$\,cm$^{-3}$, $T=100$\,K and A$_{\rm V}=2$\,mag) during 10$^6$\,yr. The second step is the \textit{pre-stellar phase}, where the cloud contracts for $\sim$5.3$\times10^6$\,yr at a fixed temperature of 10 K until it reaches a final density following a free-fall collapse parametrisation \citep{rawlings1992, holdship2017}. For the warm component, the final density is n$_{\rm H}=5\times10^{8}$\,cm$^{-3}$ \citep{quenard2018a} while for the cold component we set it to n$_{\rm H}=6\times10^{6}$\,cm$^{-3}$, in agreement with the physical structure determined for I16293's envelope \citep[][]{crimier2010}. The third phase is the \textit{warming-up or proto-stellar phase}, when the density is kept constant and the gas/grain temperatures increase from 10\,K up to either 160\,K for the warm component, or to 30\,K for the cold envelope. For all phases, we assume a constant external radiation field of G$_0=1$\,Habing, while we vary the cosmic-ray ionisation rate ($\zeta$) from 1 to 6 times the standard value ($\zeta_{\rm std}=1.3\times10^{-17}$\,s$^{-1}$). This is consistent with studies of molecular ions such as HCO$^+$ toward I16293 \citep[][]{doty2004,quenard2018b}.

\begin{figure}
	\centering
	\includegraphics[width =0.72\columnwidth]{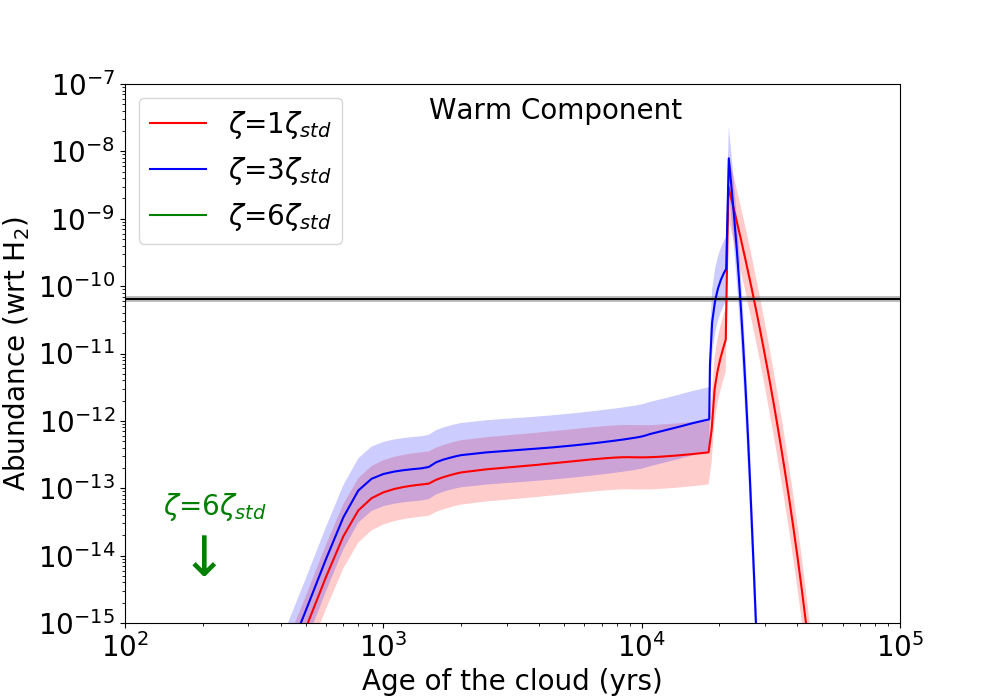}
	\includegraphics[width =0.72\columnwidth]{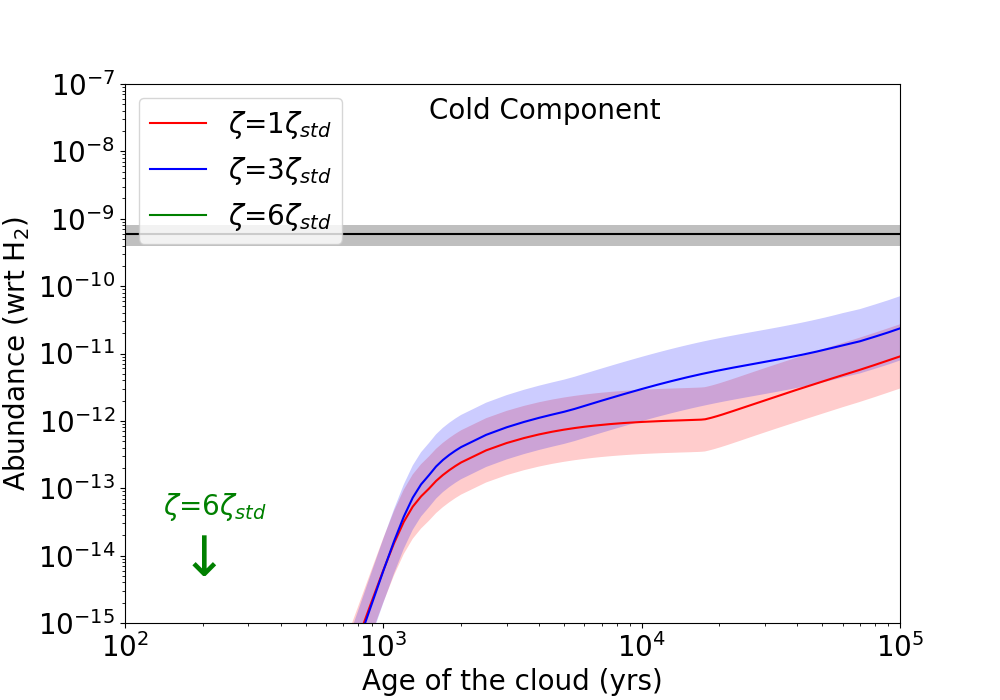}
	\vspace{-2mm}
	\caption{Observed (black line) vs modelled (colored lines) abundance of HOCH$_2$CN as a function of time (starting from the warming-up, proto-stellar phase) assuming different cosmic-ray ionisation rates, for the warm (top) and cold (bottom) components of I16293. The black area represents the error bar of the measured values. The colored areas represent the errors associated with the models, assumed to be a factor of 3.}
		\label{chemModels}
\end{figure}

Figure \ref{chemModels} presents the observed abundance of HOCH$_2$CN compared to the predicted values obtained for the warm and cold components varying $\zeta$. For the warm component (upper panel), the sudden increase of HOCH$_2$CN at time-scales $\sim$2$\times$10$^4$ yrs is due to the thermal evaporation of HOCH$_2$CN once the dust temperature ($T_{\rm dust}$) exceeds its binding energy (note that HOCH$_2$CN in our model is formed only on dust grains; see above). Despite the inherent uncertainties, the observed HOCH$_2$CN abundance seems to match the modelled abundance for $\zeta_{\rm std}$ and $\zeta=3\times\zeta_{\rm std}$ at the dynamical age of I16293B of $\sim$1-2$\times$10$^4$ yrs \citep[e.g.][and references therein]{quenard2018b}. At longer time-scales, HOCH$_2$CN is efficiently destroyed in the gas phase by ion-neutral reactions with H$_3$$^+$, HCO$^+$, and H$_3$O$^+$. For the cold component (lower panel, Figure \ref{chemModels}), since the $T_{\rm dust}$=30\,K is lower than the binding energy of HOCH$_2$CN (6980\,K), HOCH$_2$CN is non-thermally desorbed from dust grains mostly by cosmic-ray induced secondary UV-photons. The observed abundance cannot be reproduced by any model. This suggests that the network is missing key gas phase formation routes of HOCH$_2$CN. 

We finally note that the predicted abundances of the photo-products CHOCN and CH$_2$CNH are $\leq$10$^{-14}$, consistent with the upper limits inferred toward I16293B. 


In summary, the detection of HOCH$_2$CN toward I16293B indicates that this important pre-biotic molecule can be synthesized in the ISM to be incorporated into Solar-system objects such as comets and asteroids. Our modelling shows that the chemical network of HOCH$_2$CN might not be complete, and thus further experimental and theoretical work is needed to constrain its gas-phase chemistry.

\section{Acknowledgements}
This paper makes use of the following ALMA data: ADS/JAO.ALMA $\#$2011.0.00007.SV, $\#$2012.1.00712.S, $\#$2013.1.00061.S, $\#$2013.1.00352.S, and $\#$2015.1.01193.S. ALMA is a partnership of ESO (representing its member states), NSF (USA) and NINS (Japan), together with NRC (Canada), MOST and ASIAA (Taiwan), and KASI (Republic of Korea), in cooperation with the Republic of Chile. The Joint ALMA Observatory is operated by ESO, AUI/NRAO and NAOJ. We thank the anonymous referee for her/his constructive comments that contributed to improve the manuscript. S.Z. acknowledges support through a Principal's studentship funded by Queen Mary University of London. D.Q.acknowledges support received from the STFC through an Ernest Rutherford Grant (grant number ST/M004139). I.J.-S. acknowledges partial support by the MINECO and FEDER funding under grants ESP2015-65597-C4-1 and ESP2017-86582-C4-1-R. V.M.R. has received funding from the European Union's H2020 research and innovation programme under the Marie Sk\l{}odowska-Curie grant agreement No 664931. L.T. acknowledges partial support from the Italian Ministero dell$^\prime$Istruzione, Universit\'a e Ricerca through the grant Progetti Premiali 2012-iALMA (CUP C52I13000140001), by the Deutsche Forschungs-gemeinschaft (DFG, German Research Foundation)-Ref no. FOR 2634/1 TE 1024/1-1, and by the DFG cluster of excellence Origin and Structure of the Universe. R.M.D. acknowledges support by an award from the Simons Foundation (SCOL\#321183, KO). J.M.-P. acknowledges partial support by the MINECO and FEDER funding under grants ESP2015-65597-C4-1 and ESP2017-86582-C4-1-R. 



\bibliographystyle{mnras}
\bibliography{Detection_of_hydroxyacetonitrile_in_IRAS16293.bbl} 







\bsp	
\label{lastpage}
\end{document}